\documentclass{article}

\usepackage{arxiv}

\usepackage[utf8]{inputenc} 
\usepackage[T1]{fontenc}    
\usepackage{hyperref}       
\usepackage{url}            
\usepackage{booktabs}       
\usepackage{amsfonts}       
\usepackage{nicefrac}       
\usepackage{lipsum}		
\usepackage[pdftex]{graphicx}
\usepackage[square,sort,comma,numbers]{natbib}
\usepackage{doi}
\usepackage{amsmath}

\title{A Study on the relationship between the geometrical shapes and the biometrical acoustic characteristics of human ear canal}

\author{ \href{https://orcid.org/0000-0002-1801-633X}{\includegraphics[scale=0.06]{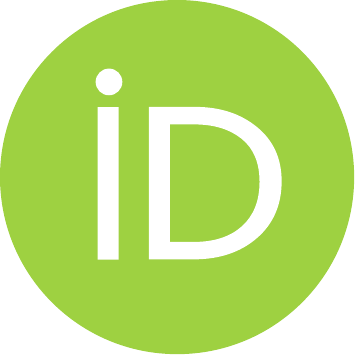}\hspace{1mm}Riki Kimura} \\
	Department of Electrical, Electronics and Information Engineering\\
	Nagaoka University of Technology\\
	Nagaoka, Japan \\
	\texttt{s215023@stn.nagaokaut.ac.jp} \\
	\And
	Shunsuke Tanaka \\
	i-TEC inc\\
	Nagaoka, Japan \\
	\And
	\href{https://orcid.org/0000-0001-7781-3876}{\includegraphics[scale=0.06]{orcid.pdf}\hspace{1mm}Naoki Wakui} \\
	Department of Electrical and Electronic Systems Engineering\\
	National Institute of Technology, Nagaoka College\\
	Nagaoka, Japan \\
	\And
	\href{https://orcid.org/0000-0001-9648-5193}{\includegraphics[scale=0.06]{orcid.pdf}\hspace{1mm}Naoki Kodama} \\
    Institute for Human Movement and Medical Sciences\\
	Niigata University of Health and Welfare\\
	Niigata, Japan \\
	\And
    Shohei Yano \\
    Department of Electrical and Electronic Systems Engineering\\
	National Institute of Technology, Nagaoka College\\
	Nagaoka, Japan \\
}

\renewcommand{\shorttitle}{Relationship between ear canal shape and acoustic characteristics}

\hypersetup{
pdftitle={A Study on the relationship between the geometrical shapes and the biometrical acoustic characteristics of human ear canal},
pdfsubject={},
pdfauthor={Riki Kimura},
pdfkeywords={},
}

\begin{document}
\maketitle

\begin{abstract}
	Ear acoustic authentication is a new biometrics method and it utilizes the differences in acoustic characteristics of the ear canal between users. However, there have been few reports on the factors that cause differences in the acoustic characteristics. We investigate the relationship between ear canal shapes and acoustic characteristics in terms of user-to-user similarity. We used magnetic resonance imaging (MRI) to measure ear canal geometry. As a result, the correlation coefficient between shape similarity and acoustic characteristic similarity is higher than 0.7 and the coefficient of determination is higher than 0.5. This suggests that the difference in the shape of the ear canal is one of the important factors.
\end{abstract}

\keywords{Biometrics \and MRI \and Ear acoustic authentication \and 3D}

\section{Introduction}
\begin{figure}[tb]
	\centering
	\includegraphics[width=0.75\linewidth]{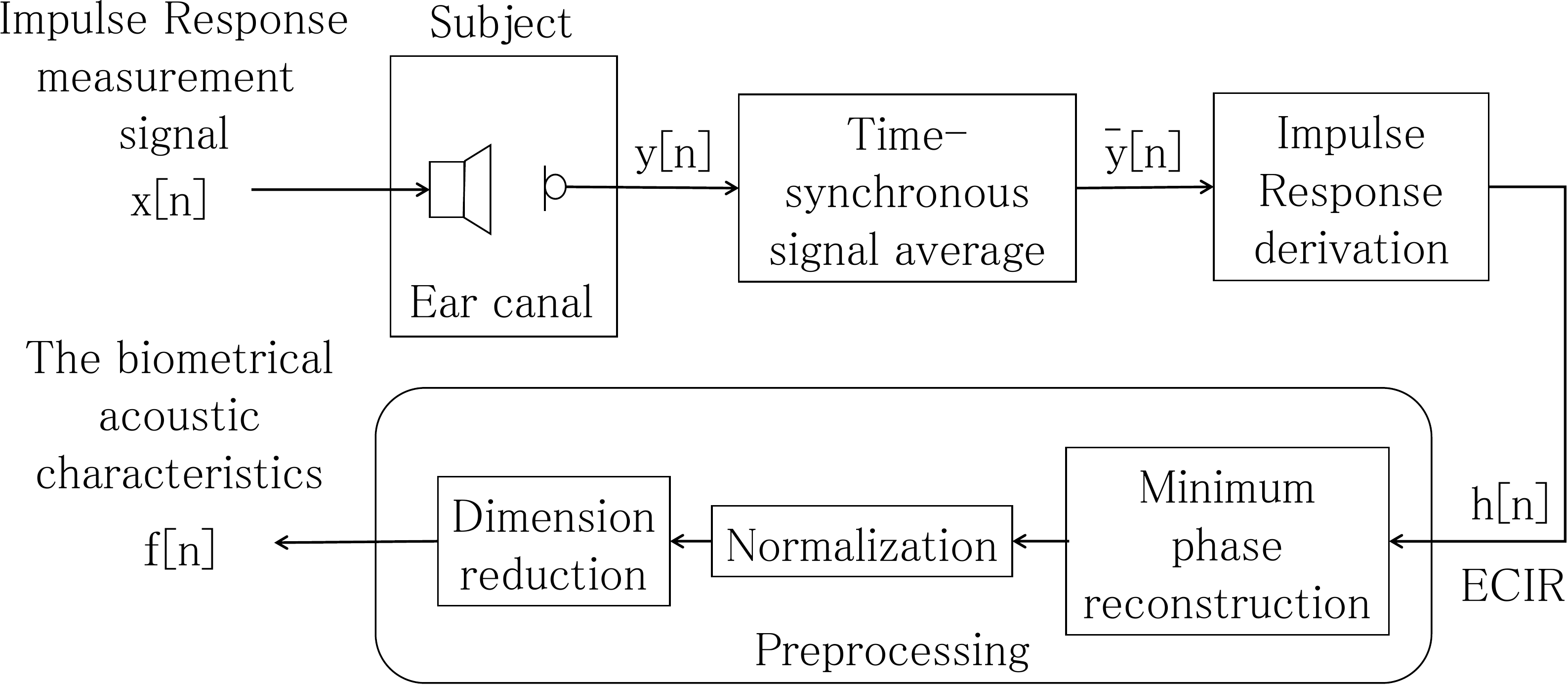}
	\caption{The flow of measuring ear acoustic feature.}
	\label{fig:sokuteikei}
\end{figure}

In recent years, biometric authentication has been attracting increasing attention as a means of personal authentication that takes advantage of the fact that biometric characteristics such as face and fingerprints vary from person to person.
Biometric authentication has been actively introduced in recent years because of its lower risk of compromise and theft compared to passwords and physical keys\cite{Mizoguchi2010, Imaoka2010, Koshinaka2015}.
Typical examples are fingerprint authentication using fingerprints and face recognition using faces.
It is used for logging in to information terminals such as smartphones and for identity verification when making online payments, and has become a familiar part of our daily lives.
However, many systems require special devices for authentication or authentication actions such as pressing a fingerprint against a sensor.
In addition, since authentication is performed at the start of the service, it is difficult to detect "spoofing," in which a user switches during the service.
Therefore, we proposed an ear acoustic authentication that can solve these problems\cite{Yano2015,Yano2017,Yasuhara2019}.
In ear acoustic authentication, the feature measurement is realized in the following flow(Figure \ref{fig:sokuteikei}).
First, the user puts on earphones with a built-in microphone.
Next, measurements of the acoustic characteristics of the ear canal are performed using earphones.
Then, features are derived from the acoustic characteristics.
The authentication is performed using a discriminator based on the differences for each user included in the features.
%
However, the mechanism that causes individual differences in the acoustic properties of the ear canal is not fully understood.
In the field of sound image localization technology, which creates a virtual sound arrival direction for earphones and headphones, it has been pointed out that differences in the shape of the outer ear and ear canal of each individual result in differences in the way people hear\cite{Moller1992, onzou, Bulauert1997}.
Therefore, we expected that the acoustic characteristics of the ear canal used for ear acoustic authentication would also vary among individuals depending on the shape of the ear canal.
Our preliminary experiments have shown that the accuracy of ear acoustic authentication is reduced when identifying twins.
It is also known that in face recognition using the three-dimensional shape of the face, the accuracy of recognition is reduced when identifying twins\cite{kaoninshou}.
From these results, we hypothesized that similar physical characteristics between the twins reduced the differences in the acoustic characteristics of the ear canal.
In this paper, magnetic resonance imaging (MRI) is used to acquire ear canal shape of four subjects, including a pair of twins.
We also propose a method for calculating the inter-subject similarity of ear canal shape using the coordinates of the center of gravity of the ear canal shape, and use it to calculate the inter-subject similarity of ear canal shape.
Next, the acoustic properties of the ear canal of the same subject are measured.
The relationship between ear canal shape and acoustic characteristics is investigated by comparing the similarity of ear canal shape and acoustic properties between subjects.
The above will allow us to investigate whether the shape of the ear canal is the cause of the differences in the acoustic characteristics of the ear canal in each person.

\section{Characteristics of the ear canal}
\subsection{Shape of the ear canal}
\begin{figure}[tb]
	\centering
	\includegraphics[width=0.8\linewidth]{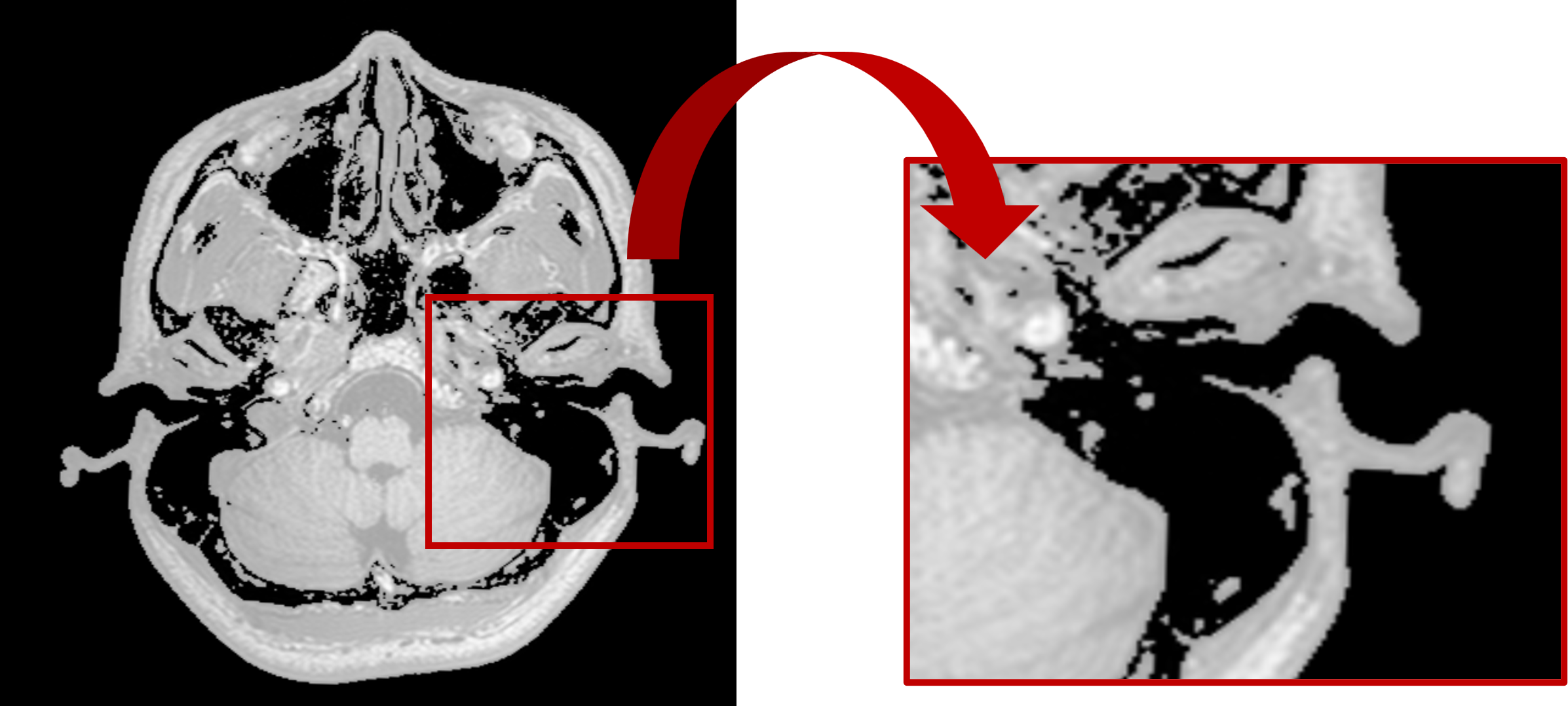}
	\caption{An example of MRI image.}
	\label{sunpou-mri}
\end{figure}

The ear canal is the name of the organ that is part of the outer ear and runs from the entrance of the hole to the eardrum.
We use MRI to obtain tomographic images of the internal structures of the human body, including the shape of the external auditory canal and surrounding organs.
MRI is noninvasive and can obtain cross-sectional, coronal, and sagittal images, as well as tomographic images in any plane.
Then, by integrating many tomograms, the three-dimensional shape of the ear canal can be measured.
Figure \ref{sunpou-mri} shows an example of MRI image.
It is evident that the shape of the external auditory canal and surrounding organs can be obtained.
In the next section, we define the thin slice shape center function as a shape property of the ear canal.
%

\subsection{Acoustic characteristics of the ear canal}
\begin{figure}[tb]
	\centering
	\includegraphics[width=0.6\linewidth]{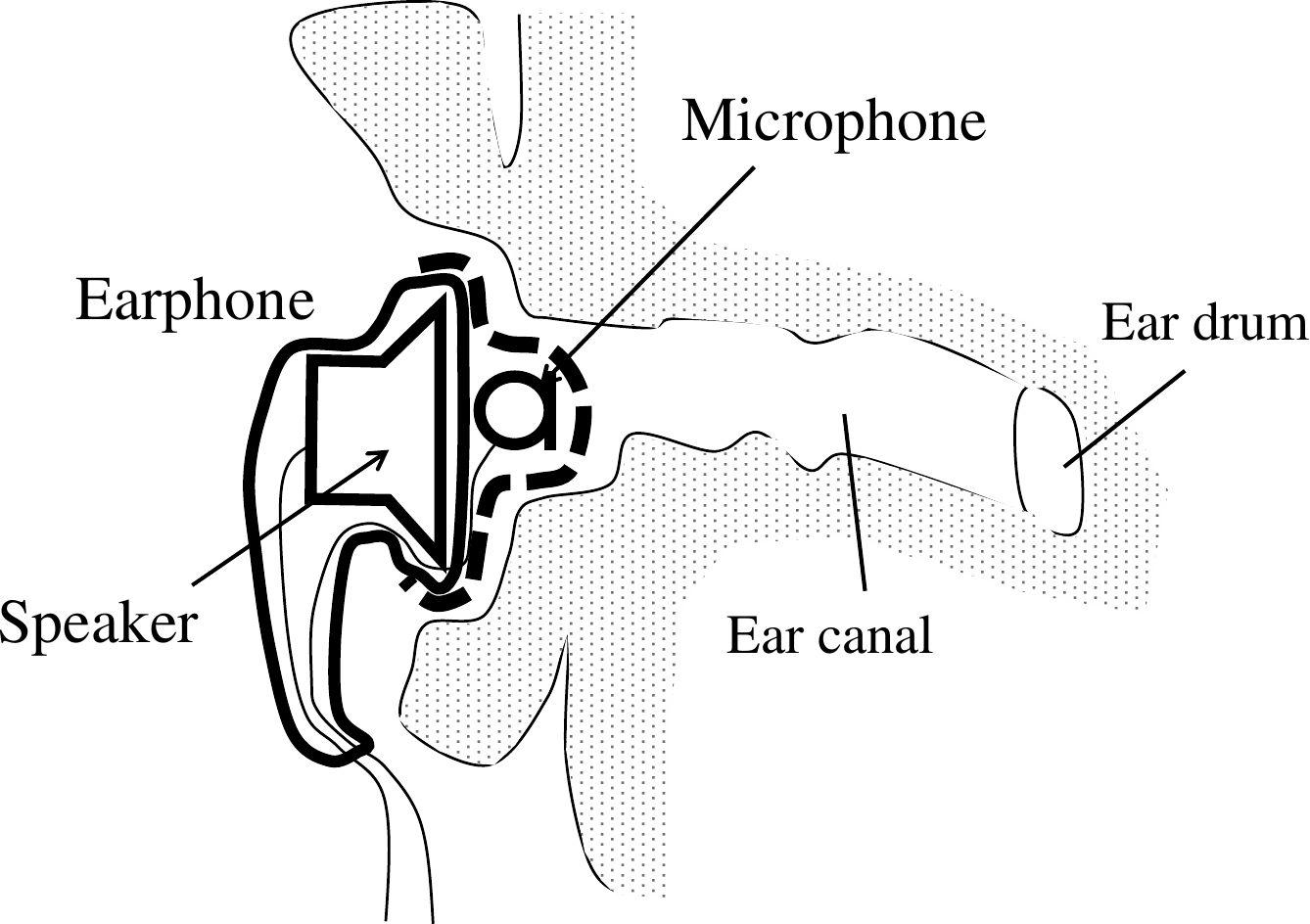}
	\caption{Sketch of outer ear with measurement device (earphone with microphone) for identification of ear canal acoustics\cite{yasuhara2022}.}
	\label{fig:eaasystem}
\end{figure}
The acoustic characteristics of the ear canal are acquired by wearing earphones with a small built-in microphone in the ear, as shown in Figure \ref{fig:eaasystem}.
They can be obtained by impulse response measurement using the Time Stretched Pulse (TSP) signal and Maximum Length Sequence (MLS) signal.
This includes the electro-acoustic conversion characteristics of the earphone, the acousto-electrical conversion characteristics of the microphone, and the characteristics dependent on the positional relationship between the microphone and the earphone, in addition to the acoustic characteristics of the ear canal.
In this paper, we define ear canal acoustic characteristics as characteristics related to signal transmission between an earphone and a microphone built into the earphone, which are obtained by impulse response measurement methods.

\section{Similarity of ear canal shape between subjects}
\subsection{Measurement of ear canal shape}\label{keijyousokutei}
Ear canal shape measurements were performed on a total of four subjects, including one pair of twins.
Subjects were males aged from 10 to 40 years old.
This study was approved by the Ethical Review Committee of Niigata University of Health and Welfare (approval number: 18343-191209).
All subjects gave informed consent to participate in this study.
Measurements were performed at Niigata University of Health and Welfare, using a 3T-MRI system (Canon, Vantage Galan).
A high-resolution magnetization-prepared rapid-gradient-echo (MP-RAGE) sequence of T1-weighted images was used for the head images, with repetition time (TR) = 5.8 msec, echo time (TE) = 2.7 msec, reversal time (TI) = 900 msec , Flip angle (FA) = 9°, number of matrices = 256 × 256, field of view (FOV) = 23 × 23 mm, and slice thickness = 1.2 mm.
The measured data was exported in Digital Imaging and Communications in Medicine (DICOM) format.
DICOM data was reconstructed and converted to Standard Triangulated Language (STL) format, one of the file formats for 3D CAD software.
Centro de Tecnologia da Informação Renato Archer (CTI) InVesalius 3 was used for the conversion.
After conversion, the spatial shape of the interior of the ear canal portion was extracted using 3D CAD software (AutoDesk Meshmixer).
It is known that MRI can usually only measure the geometry up to approximately 10 mm from the ear canal opening.
This is because the soft tissue on the surface of the ear canal is thinner at the deeper end, making it difficult to separate the ear canal from the mastoid process of the temporal bone on an imaging image.
However, in the four subjects in this experiment, we were able to obtain the shape of about one-third of the entire ear canal from the ear canal opening.
In this range, the shape and acoustic characteristics are compared.
In particular, the shape change near the aperture edge is considered to be important for the acoustic characteristics because of the free field near the aperture edge.
In STL format, the 3D shape is represented as a set of triangles, consisting of the vertex coordinates and normal vectors of the triangles.
The points of center of gravity $(x_i,y_i,z_i)$ of the triangles comprising the internal ear canal shape were derived and used in later processes.
Note that $i$ is the serial number of each triangle.
It consists of about 3,000 triangles in each ear.

\subsection{The ear canal thin slice shape center function}
\begin{figure}[tb]
	\centering
	\includegraphics[width=1\linewidth]{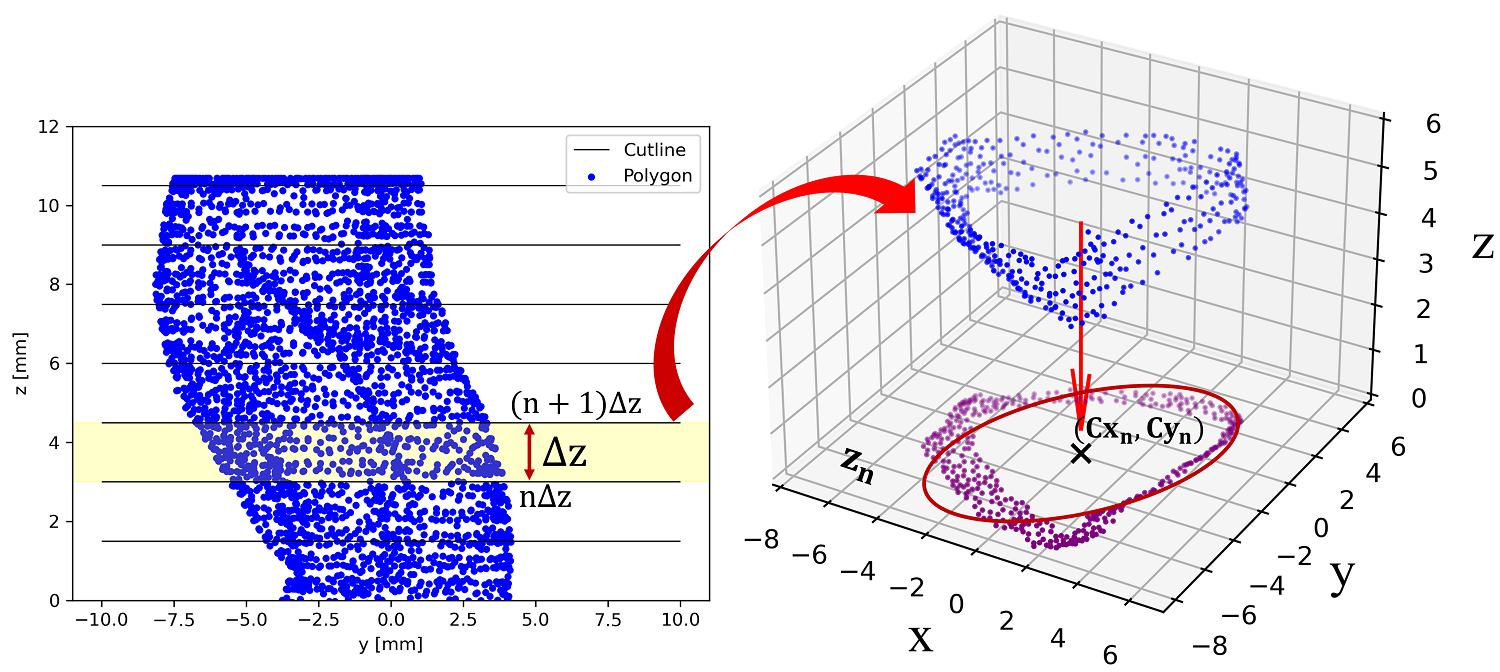}
	\caption{Derivation process of the ear canal thin slice shape center function.}
	\label{slicemethod}
\end{figure}
The ear canal thin slice shape center function is computed as follows\cite{KimuraAES};
The ear is divided equally on the axis z from the ear entrance to the tympanic membrane, and elliptic approximation is performed on the segmented surfaces.
Derive the ear canal thin slice shape center function $E_c$ from the approximate ellipse center coordinates.
In the derivation of $E_c$, first, from the set of triangular centers of gravity $\bf S_n$ = $\{(x_n, y_n)|n\Delta z < z \le (n+1)\Delta z, n=0\sim N \}$ that compose the ear canal shape existing in the division range We obtain the elliptic center coordinates $({cx}_{n}, {cy}_{n})$ by elliptic approximation.
Note that $\Delta z$ is the segment width that divides the ear canal shape in the z-axis direction.
The subscript $N$ is the number of the split face viewed from the entrance side of the ear, and its maximum number is $N$.
For the elliptic approximation, we used the direct elliptic eigenfitting algorithm proposed by Pilu et al\cite{Pilu1996}.
Equation \ref{eq:Ed} shows the ear canal thin slice shape center function $Ec[n]$.
\begin{equation}\label{eq:Ed}
	Ec[n] = (x_n, y_n) = (cx_n - cx_0, cy_n - cy_0) 
\end{equation}
$({cx}_{n},{cy}_{n})$ denotes the coordinates of the elliptic center when the set $\bf S_n$ is mapped to the $z_n$ plane perpendicular to the z axis and elliptic approximation is performed on the $z_n$ plane (Figure \ref{slicemethod}).
Since the origin coordinates vary depending on the measurement conditions of the MRI system The origin coordinates are corrected by subtracting $(cx_0, cy_0)$ at the $n=0$th ellipse center coordinates at the ear entrance.
The ear canal thin slice shape center function $Ec[n]$ represents the characteristics of each subject's ear canal shape.

\subsection{Calculation method of similarity of ear canal shape}
The shape similarity between subjects $\Phi_{s}$ are calculated from each subject's ear canal thin slice shape center function $Ec[n]$.
Cosine similarity is used to calculate shape similarity between subjects.
Assume that the subjects are $A,B$, and let $Ec_{A}, Ec_{B}$ be the thin slice shape center functions of the ear canal, respectively.
As a preprocessing step for the calculation of the cosine similarity, $\phi,\theta$ correction process is performed to correct the angle $\phi$ and rotation $\theta$ between the data.
Equation \ref{kakudohosei} shows the ear canal thin slice shape center function $Ec_{A,0},Ec_{B,\theta}$ when subject $B$ is rotated by $\theta$ with subject $A$ as the reference $\theta =0$.
In this study, the angle $\phi$ was assumed to be based on the head fixture at the time of MRI measurement, and $\phi$ correction processing was not performed.
%
\begin{eqnarray}
\begin{cases}
Ec_{A,0}[n] = \begin{bmatrix}
            	1 & 0 \\
            	0 & 1
            	\end{bmatrix}
	\begin{bmatrix}
	x_{An} \\
	y_{An}
	\end{bmatrix}\\
Ec_{B,\theta}[n] = \begin{bmatrix}
\cos\theta & -\sin\theta\\
\sin\theta & \cos\theta
\end{bmatrix}
\begin{bmatrix}
x_{Bn}\\
y_{Bn}
\end{bmatrix} 
\end{cases}\label{kakudohosei}
\end{eqnarray}
With subject A's $\theta=0$ as the reference The center function of the thin slice shape of the external auditory canal of subject B is calculated as $0 \leq \theta < 2\pi$ at the rotation angle of the xy-plane $0 \leq \theta < 2\pi$ The maximum value of the shape similarity ${{\Phi}{s}}_{A,B}$ between subjects is shown in the equation \ref{coseq}.
\begin{eqnarray}\label{coseq}
    \Phi{s}_{A,B}=\max_{\theta} \left(\frac{1}{N}\sum_{n=1}^{N}\frac{Ec_{A,0}[n]\cdot Ec_{B,\theta}[n]}{|Ec_{A,0}[n]||Ec_{B,\theta}[n]|}\right)
\end{eqnarray}

\subsection{Experimental results of ear canal shape similarity}
The shape similarity between subjects of the  ear canal shape of the right ear of one pair of twins ($Twins_A, Twins_B$) and two other subjects ($User_A, User_B$), for a total of four subjects, was calculated.
In the derivation of the ear canal thin slice shape center function, the optimal value of the division width $\Delta z$ that divides the ear canal shape in the z-axis direction has been considered in the past\cite{smasys}.
In this study, the ear canal thin slice shape center function was derived with $\Delta z = 0.1 mm$.
Table \ref{keizyouresult-crop} shows the results of the shape similarity calculations between subjects.
%
\begin{table}[tb]
\caption{The ear canal shape similarity between subjects (Right ear).)}
\centering
	\label{keizyouresult-crop}
	{
\begin{tabular}{c|c|c|c|c} 
\hline
\footnotesize
      &  $TwinsA$ & $ TwinsB$ & $UserC$  & $UserD$ \\\hline 
$TwinsA$ &	-  & $0.947$ & $0.870$ & $0.871$   \\
$TwinsB$ &	$0.947$ & - & $0.837$ & $0.826$   \\
$UserC$ & $0.870$ & $0.837$ & - & $0.759$   \\
$UserD$ & $0.871$ & $0.826$ & $0.759$ & -  \\\hline 
\end{tabular}
}

\end{table}
\begin{figure}[tb]
	\centering
	\includegraphics[width=0.9\linewidth]{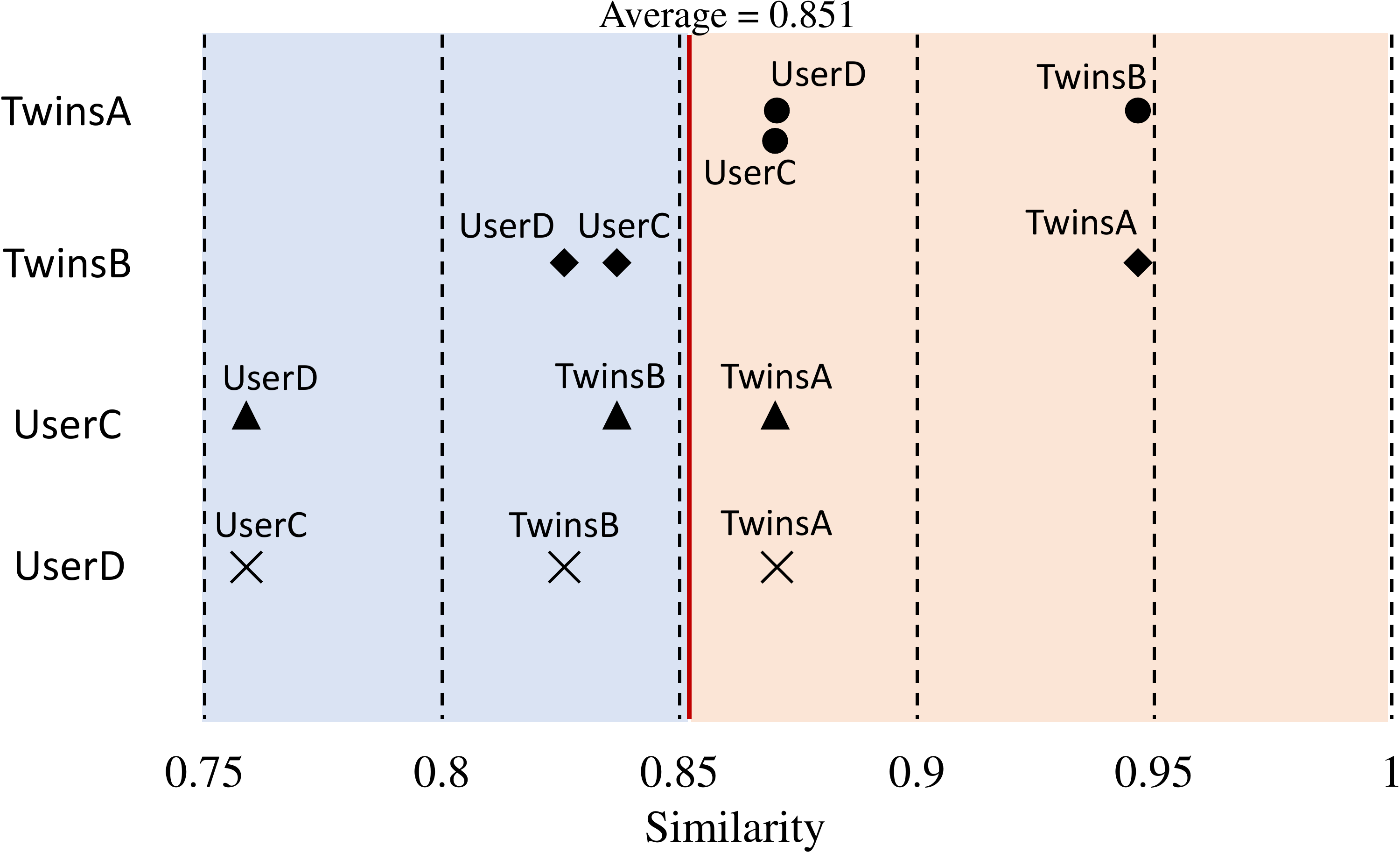}
	\caption{Comparison of the ear canal shape similarity between subjects.}
	\label{keizyouresult-fig}
\end{figure}
The relationship of the ear canal shape similarity between subjects is also illustrated in Figure \ref{keizyouresult-fig}.
The similarity of the ear canal shape between twins $\Phi{s}_{TwinA,TwinB}=0.947$, which is higher than the others.
%
%
\section{Similarity of ear canal acoustic characteristics between subjects}\label{onkyosokutei}
\subsection{Measurement of ear canal acoustic characteristics}\label{onkyosokutei_maeshori}
\begin{figure}[tb]
	\centering
	\includegraphics[width=0.75\linewidth]{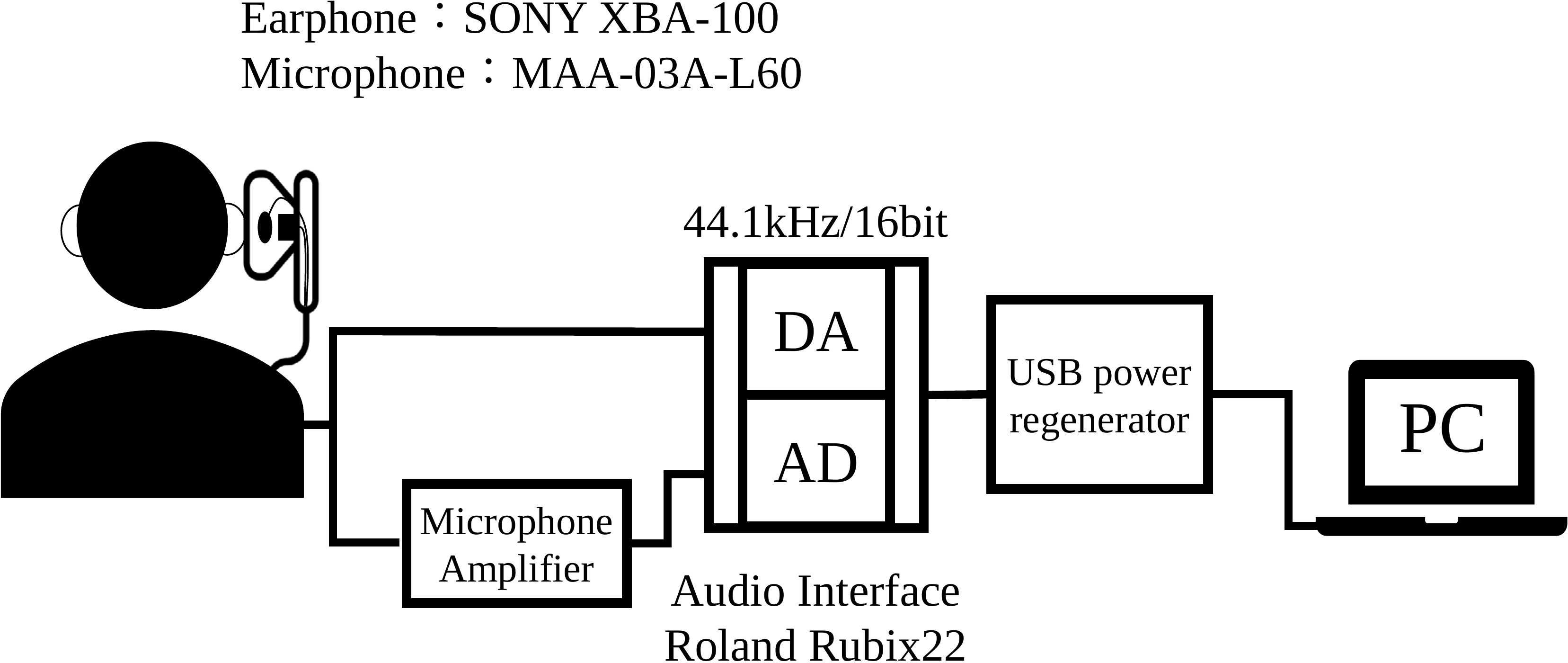}
	\caption{The measurement system for acoustic characteristics of the ear canal.}
	\label{fig:sokuenv}
\end{figure}

\begin{figure}[tb]
	\centering
	\includegraphics[width=0.6\linewidth]{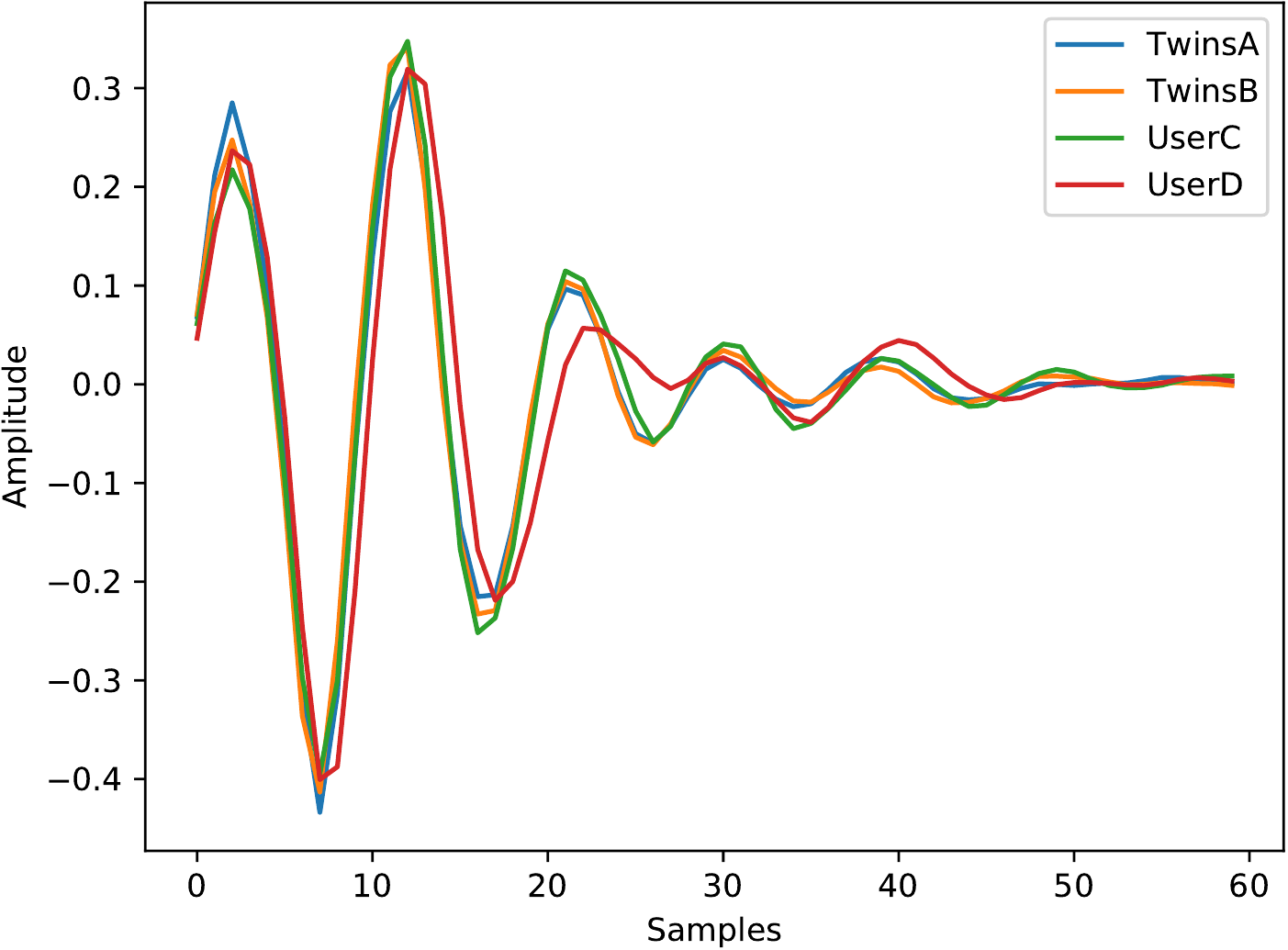}
	\caption{Comparison of time series waveforms for each subject (first 60 samples)}
	\label{fig:Time-crop}
\end{figure}

\begin{figure}[tb]
	\centering
	\includegraphics[width=0.6\linewidth]{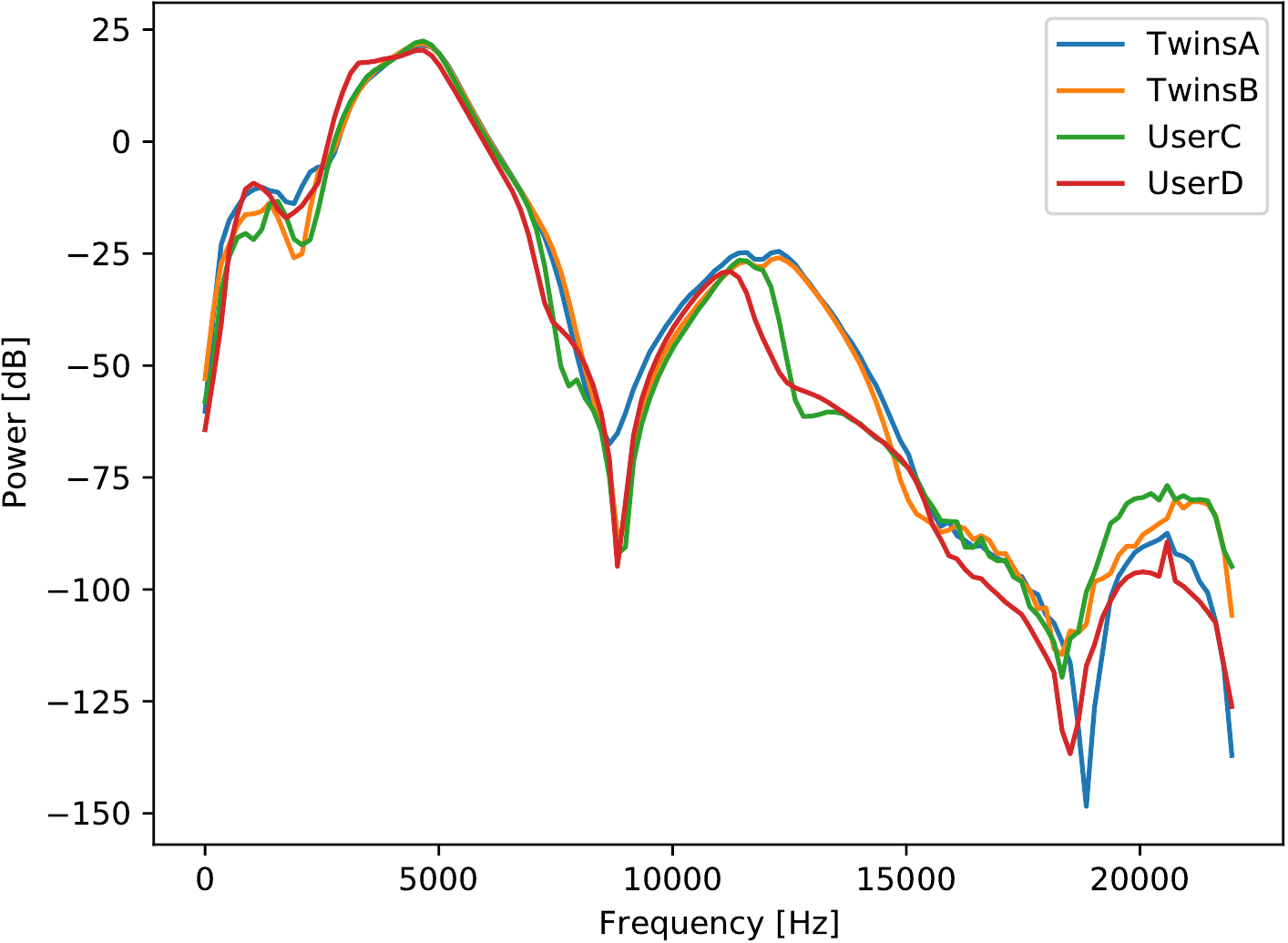}
	\caption{Comparison of power spectrum for each subject}
	\label{fig:PS-crop}
\end{figure}

The ear canal acoustic characteristics were measured 10 times per subject for the same subjects as in the \ref{keijyousokutei} section.
The measurement system is shown in Figure \ref{fig:sokuenv}.
%
The ear canal acoustic characteristics are derived using the same method as for ear acoustic authentication\cite{yasuhara2022}.
Sealed canal-type earphones (SONY XBA-100) were used as earphones.
A small microphone (STAR MICRONICS MAA-03A-L60) is placed inside the earphone.
The measurement signal reproduced from the earphone is reflected, diffracted, or interfered with in the subject's ear canal and captured by a small microphone, making it possible to acquire the acoustic characteristics of the subject's ear canal.
The sound signals were amplified using a microphone amplifier (ACO TYPE6226).
We used MLS signal (signal length $2^{16}-1$) as the measurement signal, with a sound pressure level of about 60 dB(A) at the ear and five synchronous additions.
After the impulse response derivation process, the signal was cut out before the rise time of the signal and the minimum phase was processed by Hilbert transform.
A butterworth bandpass filtering process (100 Hz$\sim$22 kHz) was performed, followed by a normalization process in which the power of the entire signal was set to 1. 
An audio interface (Roland Rubix22) was used for playback of the measurement signals and recording of the sound signals.
The sound signals are stored as 44.1 kHz/16 bit PCM data by the AD converter built into the audio interface.
A low-noise transformer power supply (ELSOUND APS-usb) was used to transmit signals to and from the PC to prevent electrical noise and other noises originating from the USB power supply from entering the system.
%
As an example, one data is randomly extracted from each subject's measurement data, and the first 60 samples, whose time-series waveform characteristics can be well confirmed, are shown in Figure\ref{fig:Time-crop} and the power spectrum is shown in Figure\ref{fig:PS-crop}.

\subsection{Calculation method of similarity of ear canal acoustic characteristics}
Acoustic characteristics similarity is derived from each subject's ear canal acoustic characteristics.
The ear canal acoustic characteristics $f[n]$ obtained by the measurement are time series data of the minimum phase system.
In addition, the time axis is unified among the data.
This paper calculates the similarity of amplitude values at the same time as a method of evaluating the similarity of data obtained by measurement.
Cosine similarity is used as the similarity calculation method.
Assume that the subjects are $A,B$, and their respective ear canal acoustic characteristics are ${f}_{A}[n],{f}_{B}[n]$.
The cosine similarity $\Phi{a}_{A,B}$ between ${f}_{A}[n],{f}_{B}[n]$ is shown in equation \ref{cosimp}.
%
\begin{eqnarray}
\Phi{a}_{A,B}
= \frac{1}{N}\sum_{n=0}^{N}\frac{{f}_{A}[n]\cdot{f}_{B}[n]}{|{f}_{A}[n]||{f}_{B}[n]|}\label{cosimp}
\end{eqnarray}
\subsection{Experimental results of ear canal acoustic characteristics similarity}\label{onkyouruizi}
The similarity of acoustic characteristics between subjects was calculated.
We used the right ear canal acoustic characteristics of a total of four subjects, including one pair of twins, who were the same subjects as the \label{keizyouzikken} chapters.
%
Table \ref{onkyoresult-crop} shows the similarity of the ear canal acoustic characteristics between the subjects.
%
\begin{table}[tb]
\caption{Acoustic characteristic similarity between subjects (Right ear).}
\centering
	\label{onkyoresult-crop}
	{
\begin{tabular}{c|c|c|c|c} 
\hline
\footnotesize
      &  $TwinsA$ & $ TwinsB$ & $UserC$  & $UserD$ \\\hline 
$TwinsA$ &	-  & $0.514\pm0.001$ & $0.402\pm0.001$ & $0.343\pm0.004$   \\
$TwinsB$ &	$0.514\pm0.001$ & - & $0.478\pm0.003$ & $0.364\pm0.001$   \\
$UserC$ & $0.402\pm0.001$ & $0.837$ & - & $0.295\pm0.001$   \\
$UserD$ & $0.343\pm0.004$ & $0.364\pm0.001$ & $0.295\pm0.001$ & -  \\\hline 
\end{tabular}
}
\\\rightline{(Average $\pm$ Standard deviation)\hspace{40pt}}
\end{table}

\begin{figure}[tb]
	\centering
	\includegraphics[width=0.9\linewidth]{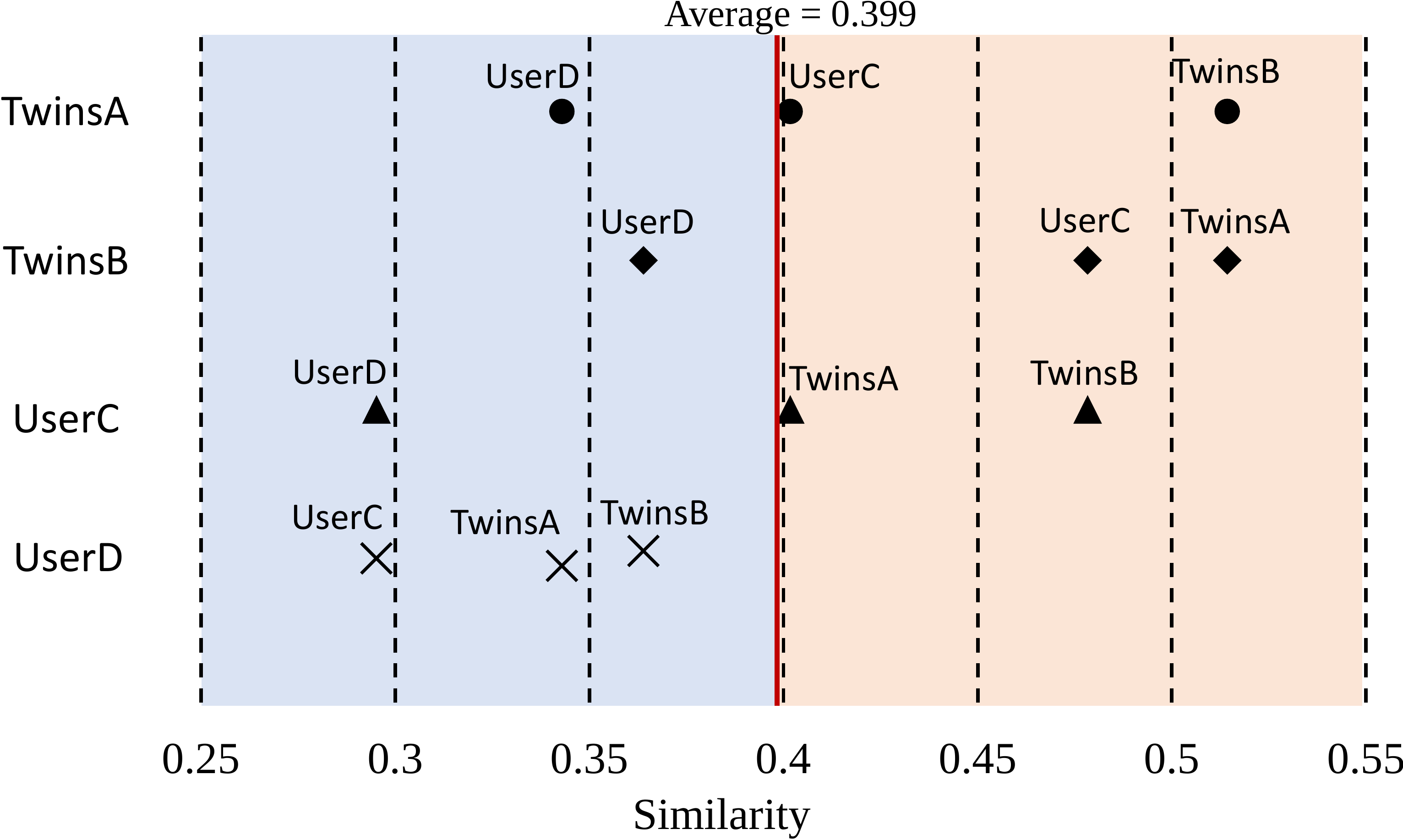}
	\caption{Comparison of acoustic characteristic similarity of the ear canal between subjects.}
	\label{onkyoresult-fig}
\end{figure}

The relationship of the ear canal acoustic characteristic similarity between subjects is also illustrated in Figure \ref{onkyoresult-fig}.
The similarity of ear canal acoustic characteristics between the twins is higher than that the others.
%
\section{Evaluation of similarity correlation between shape and acoustic characteristics}
\begin{figure}[tb]
	\centering
	\includegraphics[width=0.75\linewidth]{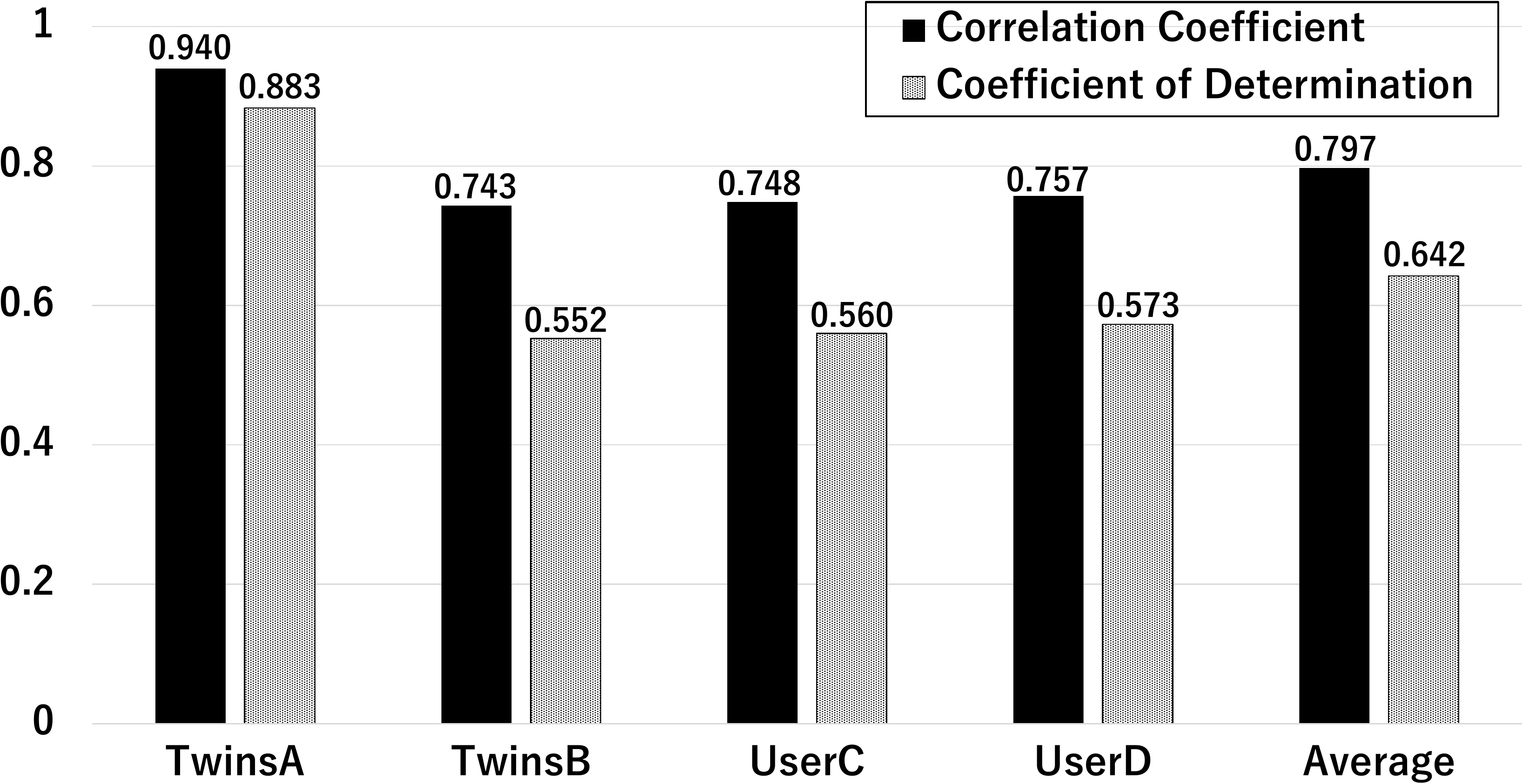}
	\caption{Correlation coefficient and coefficient of determination for each subject.}
	\label{barcor-crop}
\end{figure}

Introduce equation \ref{ASsoukanf}, which is a shape-acoustic correlation function that uses the acoustic characteristics similarity of the ear canal and the shape of the ear canal similarity as variables.
In equation \ref{ASsoukanf}, n denotes the index of subjects ($n=1\sim M$) and m denotes the index of subjects to be compared ($m=1\sim M$). The $M$ denotes the number of subjects and $M=4$.
The data of the same subject were not evaluated together($n\neq m$).
Regression analysis was performed on the shape acoustic correlation function to derive the correlation coefficient $r$ and the coefficient of determination $R^2$.
%
\begin{eqnarray}
     \Phi{as}_{n} &=& f( \Phi{s}_{n,m},\Phi{a}_{n,m} )  \label{ASsoukanf}\\ 
 &m&=1\sim M , (n \neq m) \nonumber 
\end{eqnarray}
Figure \ref{barcor-crop} shows the relationship between the correlation coefficient and the coefficient of determination for each subject.
The correlation coefficient $r$ was greater than 0.7 for all subjects, and the coefficient of determination $R^2$ was greater than 0.5 for all subjects, indicating a relationship between similarity of ear canal shape and similarity of ear canal acoustic characteristics.
The correlation coefficient $r$ is an indicator that the closer it is to 1, the stronger the positive correlation. $r \geq 0.5$ is generally considered to be a high correlation.
The coefficient of determination $R^2$ is an indicator that the closer to 1, the smaller the relative residuals are and the better the fit. $R^2 \geq 0.5$ is generally regarded as high. 
%
\section{Conclusion}
Ear canal shape was measured by MRI on four subjects, including one pair of twins.
The ear canal shape similarity evaluation experiment was conducted to quantitatively evaluate the similarity of ear canal shape between subjects by introducing a thin slice shape center function for the external auditory canal.
The results quantified the ear canal shape and obtained the relationship of shape similarity between subjects.
Compared to the overall average $\hat{\Phi_{s}}=0.851$, the average value between twins was $\Phi_{s twins}=0.947$, which is about $11.2\%$ higher. The twins who participated in this experiment are considered to have similar ear canal shapes.
%
Next, we measured the acoustic characteristics of the ear canal in the same subjects, and conducted an experiment to quantitatively evaluate whether the acoustic characteristics of the ear canal were similar between subjects.
As a result, relationship of similarity of acoustic characteristics between subjects was obtained.
Compared to the overall average $\hat{\Phi_{a}}=0.399$, $\Phi_{a twin}=0.514$, which is about $28.8\%$ higher. The twins who participated in this experiment were considered to have similar ear canal acoustic characteristics.
%
The correlation coefficient $r$ and coefficient of determination $R2$ were derived from regression analysis of the relationship between the shape of the ear canal and the acoustic characteristics.

This work was supported by JSPS KAKENHI Grant Numbers JP19H04112 and JP19K22851.

\bibliography{template}
\bibliographystyle{plain}

\end{document}